\documentclass[aps,nofootinbib]{revtex4}

\usepackage{amscd,amsmath,amssymb,enumerate}

\newcommand{\p}[1]{(\ref{#1})}

\newcommand{\cN}{{\cal N}}

\newcommand{\cS}{{\cal S}}

\newcommand{\cA}{{\cal A}}

\newcommand{\bD}{{\overline D}{}}

\newcommand{\bQ}{{\overline Q}{}}
\newcommand{\bS}{{\overline S}{}}

\newcommand{\bI}{{\overline I}{}}

\newcommand{\bV}{{\overline V}{}}
\newcommand{\bLambda}{{\overline \Lambda}{}}
\newcommand{\bPhi}{{\overline \Phi}{}}

\newcommand{\bxi}{{\bar\xi}}

\newcommand{\bpsi}{{\bar\psi}{}}

\newcommand{\bv}{{\bar v}}

\newcommand{\be}{\begin{equation}}
\newcommand{\ee}{\end{equation}}
\newcommand{\bea}{\begin{eqnarray}}
\newcommand{\eea}{\end{eqnarray}}

\newcommand{\ba}{\begin{array}} \newcommand{\ea}{\end{array}}

\def\im{{\rm i\,}}

\newcommand{\nn}{\nonumber}
\def\theequation{\arabic{section}.\arabic{equation}}

\begin{document}

\title{$\cN{=}4$ supersymmetric Schwarzian with $D(1,2;\alpha)$ symmetry}
\author{Nikolay Kozyrev}
\email{nkozyrev@theor.jinr.ru}
\affiliation{Bogoliubov  Laboratory of Theoretical Physics, JINR,
141980 Dubna, Russia}
\author{Sergey Krivonos}
\email{krivonos@theor.jinr.ru}
\affiliation{Bogoliubov  Laboratory of Theoretical Physics, JINR,
141980 Dubna, Russia}

\begin{abstract}
\noindent It was recently demonstrated that super--Schwarzian derivatives can be constructed from
the Cartan forms of the super--conformal supergroups  $OSp(1|2),SU(1,1|1), OSp(3|2), SU(1,1|2)$.
Roughly speaking, the super--Schwarzian is just the component of the corresponding Cartan forms with the
lowest dimension. In this paper, we apply the same approach for superalgebra $D(1,2;\alpha)$. The minimal set of constraints we used includes: a) introducing new superspace coordinates the Cartan forms depend on, which are completely invariant with respect to the corresponding group; b) nullifying the form for dilatation. In contrast to the $SU(1,1|2)$ case, the new super--Schwarzian appears to be a $d\theta^{ia}$ component of the form for $su(2)$ automorphism.
\end{abstract}

\maketitle
\vskip 1cm
\noindent
Receipt date: 14 January, 2022

\vskip 0.5cm


\vskip 0.5cm


\newpage

\setcounter{equation}{0}
\section{Introduction}
The breaking of the conformal symmetry in the Sachdev-Ye-Kitaev model \cite{Sachdev1, Kitaev1, Sachdev2, Sachdev3}
results in an effective Lagrangian for time reparametrization which is given by the Schwarzian. The supersymmetric versions of the SYK model up to $\cN{=}2$ supersymmetry have been constructed and analyzed \cite{super1,super2,super3}.
However, the construction of  $\cN{=}3,4$ supersymmetric SYK models and associated super--Schwarzians is not straightforward,
especially in the case of $\cN{=}4$ supersymmetry.

A new approach to the construction of Schwarzians and their supersymmetric extensions has been initiated in
\cite{Gal1} and then consistently applied to $\cN{=}1,2,3,4$ supersymmetric cases in \cite{Gal2,Gal3,Gal4,Gal5}.
The cornerstone idea of this approach is based on the invariance of the bosonic Schwarzian
$\cS( t, \tau)$ defined as
\be\label{SchwDer}
\cS( t, \tau) = \frac{\dddot t}{\dot t} - \frac{3}{2} \left( \frac{\ddot t}{\dot t}\right)^2 , \;\; \dot t = \partial_\tau t,
\ee
 under $SL(2,\mathbb{R})$ M\"{o}bius transformations acting on $t[\tau]$ via
\be\label{sl2a}
t \rightarrow \frac{a t +b}{c t + d} \,.
\ee
The immediate consequence of this statement is the conclusion that the Schwarzian can be constructed in terms of
 $sl(2,\mathbb{R})$ Cartan forms - which are essentially the unique geometric invariants of the conformal group $SL(2,\mathbb{R})$. This idea was realized in \cite{Gal1}. The straightforward generalization of this approach
 to the supersymmetric cases means passing from one dimensional conformal group $SL(2,\mathbb{R})$ to its supersymmetric
 extensions - the supergroups $OSp(1|2),SU(1,1|1), OSp(3|2), SU(1,1|2)$ and $D(1,2;\alpha)$. The relevant
 super--Schwarzians must be invariant with respect to these supergroups and, therefore, should be constructed from the
corresponding Cartan forms.

While trying to construct the Cartan forms and the invariants from them, one may encounter two problems:
\begin{itemize}
	\item One has to find a way to reduce the number of independent fields parametrizing the group element,
	\item One has to understand how the invariant (super)space\footnote{We meant the super-partners of the time $\tau$ in \p{SchwDer}.} enters the game.
\end{itemize}
The approach initiated in \cite{Gal2} works perfectly in the cases of $\cN{=}0,1$ supersymmetries, but it puts unreasonably strong conditions in the cases of higher supersymmetries. In our recent paper \cite{KK} we proposed the set of constraints which perfectly reproduced all known super--Schwarzians till $\cN{=}4$ one, related to the supergroup $SU(1,1|2)$.
These constraints can be easily summarized as follows:
\begin{itemize}
	\item For the supergroup containing the super Poincar\'{e} subalgebra $\left\{ Q_i, Q_j \right\} = 2 \delta_{ij} P$
	the invariant superspace $\left\{\tau, \theta_i\right\}$ should be introduced as
	$$ \omega_P = d\tau - \im d\theta^i \theta^i, \; \omega_Q^i = d \theta^i \qquad \qquad (a)$$
	\item The unique additional constraint is
	$$\omega_D=0 \qquad \qquad (b).$$
\end{itemize}	
Here, $\omega_P, \omega_Q^i$ and $\omega_D$ are the Cartan forms for translation, super-translations and dilatation, respectively.

In this paper we are going to demonstrate that our approach works perfectly in the most complicated case - the super--Schwarzian associated with the most general $\cN{=}4$ superconformal group $D(1,2;\alpha)$ \cite{Sorba}.
Despite the simplicity of the constraints $(a)$ and $(b)$, their application is not trivial. Indeed, one may quickly see that the straightforward calculations shortly become quite complicated and rather cumbersome. That is why we decided to use the Maurer--Cartan equations, which drastically simplify the analysis.
Thus, to be able to check the basic steps we put the main formulas in the body of the paper, transferring more technical things to the Appendices A, B and C.

\setcounter{equation}0
\section{Preliminary steps: superalgebra, Cartan forms and all that}
\subsection{Superalgebra $D(1,2;\alpha)$ }
The structure of the superalgebra $D(1,2;\alpha)$ is quite simple: it contains nine bosonic generators $T_1^{AB},T^{ij},J^{ab}$ spanning three commuting sub-algebras $sl(2) \times su(2) \times su(2)$. The eight fermionic
generators $G^{A,i,a}$ transform as the doublets with respect to each of these algebras. The anti-commutator of the
fermionic generators contains all bosonic generators as
\be\label{alg1}
\left\{ G^{A,i,a}, G^{B,j,b}\right\} \sim \epsilon^{ab} \epsilon^{ij} T_1^{AB} +\alpha \epsilon^{AB} \epsilon^{ij} J^{ab} -(1+\alpha) \epsilon^{AB} \epsilon^{ab} T^{ij} .
\ee
Here, all indices can take values $1$ or $2$, and $\epsilon^{ij}$, $\epsilon^{ab}$, $\epsilon^{AB}$ are antisymmetric symbols, normalized as $\epsilon^{21}=1$.
The parameter $\alpha$ measures the balance between two $su(2)$ subalgebras. For the two values of $\alpha=0,-1$ one of the $su(2)$ sub-algebras decouples and  $D(1,2;\alpha)$ reduces to the $su(1,1|1)\times su(2)$ superalgebra. Another interesting case corresponds to $osp(4|2)$ algebra with $\alpha =-\frac{1}{2}$, when both $su(2)$ sub-algebras occur in the same way. In what follows, we exclude consideration of the cases with $\alpha=0,-1$ which can be found in \cite{KK}. Thus, we may easily divide any expressions  by $\alpha$  and/or by $\alpha+1$.

From a physical point of view, the $sl(2)$ subalgebra $T_1^{AB}$ is the conformal algebra of one-dimensional space. Therefore, it is natural to introduce the generators of translation, dilatation and conformal boosts as \cite{ikl1}
\be
P= T_1^{22}, D= - T_1^{12}, K = T_1^{11} .
\ee
Correspondingly, the supercharges divide into ordinary $Q^{ia} $ and superconformal  $S^{ia}$ ones, as
\be
Q^{ia} = - G^{2ia}, \quad S^{ia} = G^{1ia} .
\ee
The full list of the non-zero (anti)commutators can be found in the Appendix A.
\subsection{Cartan forms}
To obtain the $D(1,2,\alpha)$-invariant super--Schwarzian, we are going to use the method of nonlinear realizations, developed in \cite{coset11,coset12,coset21,coset22}. In the present case we need to construct a nonlinear realization of the superconformal group $D(1,2;\alpha)$ with the group element $g$  parameterized as
\be\label{gD12}
g= e^{\im t P} e^{\xi_{ia} Q^{ia}} e^{\psi_{ia} S^{ia}} e^{\im z K} e^{\im u D}
e^{\im v_{ij} T^{ij}} e^{\im \phi_{ab}J^{ab}}.
\ee

The Cartan forms $\Omega$ are defined in a standard way as
\be\label{CF1}
\Omega = g^{-1} d g =\im \omega_D D + \im \omega_K K +\im \omega_P P + \im \big( \omega_J \big)_{ab}J^{ab} + \im \big( \omega_T  \big)_{ij} T^{ij} + \big( \omega_Q  \big)_{ia}Q^{ia} + \big( \omega_S  \big)_{ia}S^{ia}.
\ee

The Cartan forms for the scalar generators can be easily computed \footnote{ The $su(2)$ indices are raised and lowered as $A_i = \epsilon_{ij}A^j, \; A^j=\epsilon^{ji}A_i$, where the antisymmetric tensor $\epsilon^{ij}$ satisfies $\epsilon_{ij}\epsilon^{jk} = \delta_i^k$, $\epsilon_{12}=\epsilon^{21}=1$. }
\bea\label{scalarCF}
\omega_P& = &e^{-u} \left( d t - \im d\xi_{ia} \xi^{ia} \right) \equiv e^{-u} \triangle t , \nn \\
\omega_D & = & d u - 2 z \triangle t + 2 \im \psi_{ia} d \xi^{ia}, \nn \\
\omega_K & = & e^u\left( d z + z^2 \triangle t - 2 \im z \psi_{ia} d\xi^{ia}
+ \im \psi_{ia} d\psi^{ia} -\frac{2}{3} (1+2\alpha) \psi_k^c \psi^{kb}\psi_b^j d\xi_{jc}+\frac{1}{6} \triangle t (1+2\alpha) \psi_{ia} \psi^{ib} \psi_{jb}\psi^{ja} \right).
\eea

The fermionic  and  $su(2)\times su(2)$ forms  look more complicated\footnote{We define the matrix-valued functions
$\left( e^v\right)_i^j$ and $\left( e^\phi\right){}_a^b$ in a standard way: $n$-th term in Taylor series expansion
of $e^v$ is understood as $\frac{1}{n!} v_i^{k_1}v_{k_1}^{k_2} \ldots v_{k_{n-1}}^j$. In particular, this imply
$\left( e^v\right)_i^j = \cos\sqrt{\frac{v^2}{2}} \delta_i^j + \frac{\sin \sqrt{\frac{v^2}{2}}}{\sqrt{\frac{v^2}{2}}}v_i^j, \; v^2 = v^{ij}v_{ij}$, etc. }	
\bea \label{fermionCF}
\left(\omega_Q\right)_{ia} & = & e^{-\frac{u}{2}} \left( e^v\right)_i^j \left( e^\phi\right){}_a^b\left( d\xi_{jb}-\psi_{jb} \triangle t\right), \nn \\
\left(\omega_S\right)_{ia} & = & e^{\frac{u}{2}} \left( e^v\right)_i^k \left( e^\phi\right){}_a^c\left( d\psi_{kc} -\frac{\im}{2} \psi_{jb}d\xi^{jb} \psi_{kc}+ \frac{\im}{2} \alpha \left(\psi_{jb}d\xi^j_c+\psi_{jc}d\xi^j_b\right)\psi^b_k -\frac{\im}{2}(1+\alpha) \left( \psi_{jb}d\xi^b_k+\psi_{kb}d\xi^b_j\right)\psi^j_c -\right. \nn \\
&& \left. - \frac{\im}{3} (1+2\alpha) \triangle t \psi_k^b \psi_{jb}\psi^j_c - z ( d\xi_{kc}-\triangle t \psi_{kc})\right),
\eea
and
\be\label{full_omega_T}
\left(\omega_T\right)_{km} =\epsilon_{kj}\left( e^{-v}\right)^j_{i}d \left(e^v\right){}^i_m + \left( e^v\right)^i_k \left( e^v\right)^j_m \left({\hat\omega}_T\right)_{ij}\; \mbox{and} \;
\left(\omega_J\right)_{ab} =\epsilon_{ad}\left( e^{-\phi}\right)^d_{c}d \left(e^\phi\right){}^c_b + \left( e^\phi\right){}^c_a \left( e^\phi\right){}^d_b \left({\hat\omega}_J\right)_{cd} ,
\ee
where
\bea\label{su2CAF}
\im \left({\hat\omega}_T\right)_{ij} & = & (1+\alpha) \left( - \psi_{ib} \psi^b_j \triangle t+\psi_{ia} d\xi^a_j+\psi_{ja} d \xi^a_i\right), \nn \\
\im \left({\hat\omega}_J\right)_{ab} & = & \alpha \left( \psi_{ja} \psi^j_b \triangle t - \psi_{ja} d \xi^j_b- \psi_{jb}d\xi^j_a\right).
\eea

\subsection{Constraints and their consequences}
As we already said in the Introduction, the constraints we have to impose on the Cartan forms to find a proper
super--Schwarzian read
\bea
&& \omega_P = d\tau -\im d\theta_{ia}  \theta^{ia} \equiv \triangle \tau, \quad \omega_Q^{ia} = d\theta^{ia}, \label{a} \\
&& \omega_D = 0 . \label{b}
\eea
Note that  the covariant derivatives with respect to $\tau$, $\theta_{ia}$ are
\be\label{covders}
\partial_\tau = \frac{\partial}{\partial \tau}, \;\; D^{ia} =  \frac{\partial}{\partial \theta_{ia}} + \im \theta^{ia} \frac{\partial}{\partial \tau}, \;\; \big\{ D^{ia}, D^{jb}  \big\} = 2\im \epsilon^{ij}\epsilon^{ab}\partial_\tau.
\ee
Therefore, for any superfield $\cA$ we have
\be
d \cA = \triangle \tau \; \partial_\tau \cA + d\theta^{ia}  D_{ia} \cA.
\ee
With our definitions of the Cartan forms \p{scalarCF}, \p{fermionCF}, \p{full_omega_T} the constraints \p{a} and \p{b} read
\bea
&& \omega_P = e^{-u}\big( dt -\im d\xi_{jb}\xi^{jb}  \big) = e^{-u} \triangle t = \triangle \tau , \quad
\big(\omega_Q\big)_{ia} = e^{-u/2}\big( e^v \big)_i^j \big( e^\varphi \big)_a^b \big( d\xi_{jb} - \triangle t \psi_{jb}  \big) = d\theta_{ia} \label{a1} \\
&& \omega_D = d u - 2 z \triangle t + 2 \im \psi_{ia} d \xi^{ia} =0. \label{b1}
\eea
The  constraints \p{a1} imply
\bea
&& D^{ia}t - D^{ia}\xi_{jb}\xi^{jb} =0, \;\; \dot t - \im \dot\xi_{ia}\xi^{ia} = e^u, \label{a21} \\
&& D^{jb}\xi_{ia} = e^{u/2}\big( e^{-v} \big)_i^j \big( e^{-\varphi} \big)_a^b, \;\; \psi_{ia} = e^{-u} \dot\xi_{ia},
\label{a22}
\eea
while the constraints \p{b1} are resolved by the following relations
\be\label{b2}
z = \frac{1}{2} e^{-u} \dot{u}, \quad D_{ia}u = 2 \im \psi_{jb} D_{ia} \xi^{jb} .
\ee
We observe that it is possible to covariantly express some of the group parameters in terms of other parameters and their derivatives. This a manifestation of the Inverse Higgs phenomenon \cite{ih}. Note that these constraints involve both $d\tau$ and $d\theta$ projections of the forms, unlike the constraints considered in \cite{Gal5}. This allows, in particular, to express $z$ in terms of $u$ (or $\xi^{ia}$) without putting any constraint on the forms of the superconformal generators.

As a result of \p{a22} and the following identities
$$ \epsilon^{ik} \epsilon_{jl} \left( e^v\right)_k^l = - \left(e^{-v}\right)^i_j,  \quad \mbox{and} \quad
\epsilon^{ac} \epsilon_{bd} \left( e^\phi\right)_c^d = - \left(e^{-\phi}\right)^a_b ,$$
$D^{ia}\xi_{jb}$ satisfies the relations
\be\label{a3}
D_{jb}\xi^{ia} = e^{u/2} \big( e^{v} \big)_j^i \big( e^{\varphi} \big)_b^a \;\; \Rightarrow \;\; D^{ia}\xi_{kc} D_{jb}\xi^{kc} = \delta^i_j \delta^a_b e^u, \;\; D^{kc}\xi_{jb} D_{kc}\xi^{ia} = \delta^i_j \delta^a_b e^u,
\ee
and, moreover,
\be\label{a4}
D_{kc}\xi^{ia}\, D^{kb}\xi_{ld} = \frac{1}{2}\delta_l^i \, D_{kc}\xi^{ma}D^{kb}\xi_{md}, \;\; \mbox{etc.}
\ee

Thus, we see that all our superfields - coordinates of the group element $g$ \p{gD12}  can be expressed through the
derivatives of the superfields $\xi_{ia}$, only \footnote{The superfield $t$ can be in principle found from the equations \p{a21}.}. In principle, it should be the end of the story and the technical step is to find among the
components of the surviving Cartan forms the super--Schwarzian. Unfortunately, this technical step is too involved and the direct straightforward calculations quickly become a rather cumbersome. The simplest solution we found is to use the Maurer--Cartan equations to rewrite the Cartan forms with the constraints \p{a} and \p{b} taken into account.

\setcounter{equation}0
\section{$\cN{=}4$ super--Schwarzian}
\subsection{Maurer--Cartan equations}
If the Cartan form $\Omega$ is defined as in \p{CF1}
$$
\Omega(d) = g^{-1} d g = \im \omega_D D + \im \omega_K K +\im \omega_P P + \im \big( \omega_J \big)_{ab}J^{ab} + \im \big( \omega_T  \big)_{ij} T^{ij} + \big( \omega_Q  \big)_{ia}Q^{ia} + \big( \omega_S  \big)_{ia}S^{ia},
$$
then by construction it obeys the Maurer--Cartan equation. We prefer to deal with this equation in the form used in \cite{OSP14}. There, two independent differentials $d_1$, $d_2$ were introduced, so that $d_1 d_2=d_2 d_1$ and differentials of bosons and fermions are bosons and fermions, respectively. Therefore, the relation
\be\label{MaurerCartan}
d_2\Omega_1 - d_1 \Omega_2 = \big[ \Omega_1, \Omega_2   \big], \;\; \Omega_1 = \Omega(d_1), \;\; \Omega_2 = \Omega(d_2)
\ee
turns into identity upon substitution $\Omega_i = g^{-1}d_i g$. At the same, one can substitute $\Omega_i$ as a general expansion in generators \p{CF1} and find the list of relations the individual forms satisfy. In the case of $D(1,2;\alpha)$ equation \p{MaurerCartan} can be expanded into following set of equations \footnote{Round brackets are used to denote symmetrization of indices, $A_{(ij)} = \frac{1}{2}\big( A_{ij} + A_{ji}  \big)$}
\bea\label{MCexp}
\im \big( d_2 \omega_{1P} - d_1 \omega_{2P} \big) &=& -\im \big[ \omega_{2D}\omega_{1P} - \omega_{1D}\omega_{2P}   \big] +2\big(\omega_{1Q}\big)_{ia}\big(\omega_{2Q}\big)^{ia}, \nn \\
\im \big( d_2 \omega_{1K} - d_1 \omega_{2K} \big) &=& \im \big[\omega_{2D}\omega_{1K} - \omega_{1D}\omega_{2K}   \big] +2\big(\omega_{1S}\big)_{ia}\big(\omega_{2S}\big)^{ia}, \nn \\
\im \big( d_2 \omega_{1D} - d_1 \omega_{2D} \big) &=& -2\im \big[ \omega_{2K}\omega_{1P} - \omega_{1K}\omega_{2P}  \big] + 2\big[ \big(\omega_{1Q}\big)_{ia}\big(\omega_{2S}\big)^{ia} - \big(\omega_{2Q}\big)_{ia}\big(\omega_{1S}\big)^{ia} \big],  \\
\im \big( d_2 \big( \omega_{1J}  \big)_{ab} -  d_1 \big( \omega_{2J}  \big)_{ab} \big) &=& \im \big[ \big( \omega_{1J} \big)_{ac} \big(\omega_{2J} \big)_{b}{}^c - \big(\omega_{2J} \big)_{ac} \big(\omega_{1J} \big)_{b}{}^c  \big] -2\alpha \big[ \big( \omega_{1Q}  \big)_{i(a}  \big( \omega_{2S}  \big)_{b)}^i - \big( \omega_{2Q}  \big)_{i(a}  \big( \omega_{1S}  \big)_{b)}^i  \big], \nn \\
\im \big( d_2 \big( \omega_{1T}  \big)_{ij} -  d_1 \big( \omega_{2T}  \big)_{ij} \big) &=& \im \big[ \big(\omega_{1T} \big)_{ik} \big(\omega_{2T} \big)_{j}{}^k - \big(\omega_{2T} \big)_{ik} \big(\omega_{1T} \big)_{j}{}^k  \big] +2(1+\alpha) \big[ \big( \omega_{1Q}  \big)_{(ia}  \big( \omega_{2S}  \big)_{j)}^a - \big( \omega_{2Q}  \big)_{(ia}  \big( \omega_{1S}  \big)_{j)}^a\big], \nn \\
d_2 \big( \omega_{1Q}  \big)_{ia} - d_1 \big( \omega_{2Q}  \big)_{ia} &=& -\frac{1}{2} \big[ \omega_{2D} \big( \omega_{1Q} \big)_{ia} -\omega_{1D} \big( \omega_{2Q} \big)_{ia}     \big] +\big[\big( \omega_{1J}\big)_{ab}\big( \omega_{2Q} \big)_i^b -    \big( \omega_{2J}\big)_{ab}\big( \omega_{1Q} \big)_i^b \big]+\nn \\&& + \big[ \big( \omega_{1T}\big)_{ij}\big( \omega_{2Q} \big)_a^j -    \big( \omega_{2T}\big)_{ij}\big( \omega_{1Q} \big)_a^j \big] - \big[ \omega_{1P} \big( \omega_{2S}  \big)_{ia} -  \omega_{2P} \big( \omega_{1S}  \big)_{ia} \big], \nn \\
d_2 \big( \omega_{1S}  \big)_{ia} - d_1 \big( \omega_{2S}  \big)_{ia} &=& \frac{1}{2} \big[ \omega_{2D} \big( \omega_{1S} \big)_{ia} -\omega_{1D} \big( \omega_{2S} \big)_{ia}     \big] +\big[\big( \omega_{1J}\big)_{ab}\big( \omega_{2S} \big)_i^b -    \big( \omega_{2J}\big)_{ab}\big( \omega_{1S} \big)_i^b \big]+\nn \\&& + \big[ \big( \omega_{1T}\big)_{ij}\big( \omega_{2S} \big)_a^j -    \big( \omega_{2T}\big)_{ij}\big( \omega_{1S} \big)_a^j \big] + \big[ \omega_{1K} \big( \omega_{2Q}  \big)_{ia} -  \omega_{2K} \big( \omega_{1Q}  \big)_{ia} \big]. \nn
\eea
The forms should be subjected to the conditions
\be\label{maincond}
\omega_P = \triangle \tau, \;\; \big(\omega_{Q}\big)_{ia} = d\theta_{ia}, \;\; \omega_D =0, \;\; \triangle\tau = d\tau - \im d\theta_{jb}\theta^{jb}.
\ee
To analyze the consequences of these constraints let us represent other forms in most general way as
\bea\label{strsf}
\big( \omega_S \big)_{ia} &=& \triangle\tau \Psi_{ia} + d\theta_{jb}A_{ia}{}^{jb}, \;\; \omega_K = \triangle\tau C + d\theta_{ia}\Gamma^{ia}, \nn \\
\big(\omega_J \big)_{ab} &=& \triangle\tau \big( \cS_J   \big)_{ab} + d\theta_{kc}\Sigma_{(ab)}{}^{kc}, \;\; \big(\omega_T \big)_{ij} = \triangle\tau \big( \cS_T   \big)_{ij} + d\theta_{kc}\Pi_{(ij)}{}^{kc}.
\eea
Here $\Psi_{ia}, A_{ia}{}^{jb}, C, \Gamma^{ia}, \big( \cS_J   \big)_{ab}, \Sigma_{(ab)}{}^{kc},\big( \cS_T   \big)_{ij}$ and  $\Pi_{(ij)}{}^{kc}$ are superfields that depend on $\tau, \theta_{ia}$.

The first of equations, $d\omega_P$ in \p{MCexp}, is satisfied identically  due to the condition \p{maincond}.
Indeed, the left hand side of the first equation in \p{MCexp} reads
\be\label{lhs1}
\im \big( d_2 \omega_{1P} - d_1 \omega_{2P} \big)=\im( -\im d_1\theta_{ia} d_2 \theta^{ia} +\im d_2\theta_{ia} d_1 \theta^{ia} ) =2 d_1\theta_{ia} d_2 \theta^{ia}.
\ee
Clearly, \p{lhs1} coincides with
\be
\label{rhs1}
2\big(\omega_{1Q}\big)_{ia}\big(\omega_{2Q}\big)^{ia} = 2 d_1\theta_{ia} d_2 \theta^{ia} .
\ee
The analysis of other Maurer--Cartan equations in \p{MCexp} is straightforward, but it is rather involved.
These technical calculations are presented in the Appendix C. The result of these analysis can be summarized as follows:
the Cartan forms can be expressed through the fermionic superfield $\sigma^{ia}$ as
\bea
&& \omega_P = \triangle \tau, \quad \omega_D=0,\quad \omega_K = \triangle \tau C + \im d\theta_{ia} \Psi^{ia}, \label{eq1}\\
&& \left(\omega_J\right)_{ab}  = \triangle \tau \left(\cS_J\right)_{ab}+\frac{1}{3}\left( d\theta_{ka} \sigma^k_b+
d\theta_{kb} \sigma^k_a \right),\label{eq2} \\
&& \left(\omega_T\right)_{ij}  = \triangle \tau \left(\cS_T\right)_{ij}-\frac{1}{3}\left( d\theta_{ic} \sigma^c_j+
d\theta_{jc} \sigma^c_i \right), \label{eq3}\\
&& \left(\omega_Q\right)_{ia} = d \theta_{ia}, \quad
\left(\omega_S\right)_{ia} = \triangle \tau \Psi_{ia} -d\theta_{ib}  \left(\cS_J\right)^b_{a}
-d\theta_{ka}  \left(\cS_T\right)^k_{i} , \label{eq4}
\eea
where the superfields $C, \Psi^{ia}, (\cS_T)_{ij} ,(\cS_J)_{ab}$ have the form
\bea\label{formssol}
\left( \cS_J\right)^{ab} & = & \frac{\im}{1+\alpha} \left[\frac{1}{12} \left( D^{ka}\sigma_k^b+D^{kb}\sigma_k^a\right)-
\frac{1}{9} \sigma^a_m \sigma^{mb}\right], \nn \\
\left( \cS_T\right)^{ij} & = & \frac{\im}{\alpha} \left[\frac{1}{12} \left( D^{ic}\sigma_c^j+D^{jc}\sigma_c^j\right)+
\frac{1}{9} \sigma^i_c \sigma^{jc}\right], \nn \\
\Psi_a^k & = & \frac{\im}{3 \alpha} \left[ D^{kb} \left(\cS_J\right)_{ab} +\dot{\sigma}^k_a +\frac{4}{3} \left( \cS_J\right)_a^d \sigma_d^k\right], \nn \\
C & = & \frac{1}{4}\left[ D^{ia}\Psi_{ia} + 2 \left( \cS_J\right)^2+2 \left( \cS_T\right)^2 \right].
\eea
In addition, the superfield $\sigma^{ia}$ in virtue of the same constraints \p{maincond} has to obey the following conditions
\be\label{sigmaconstr}
\frac{1}{2} \left[ D^{ia}, D^{jb}\right] \sigma_{jb} = 3 \im \dot{\sigma}^{ia}, \quad
D^{i(a} \sigma^{jb)} + D^{j(a} \sigma^{ib)} =0.
\ee
Clearly, the fermionic superfield $\sigma^{ia}$ is a candidate for the super--Schwarzian. The final step is to express $\sigma^{ia}$ in terms of $\xi^{ia}$.
\subsection{The super--Schwarzian}
To find the explicit expression for the super--Schwarzian, one should calculate the $d\theta$-projections of $\omega_T$, $\omega_J$ forms, taking into account explicit consequences of conditions \p{maincond}. Expanding  \p{maincond} into $\triangle\tau$ and $d\theta$ projections, one can find
\bea\label{maincondcons}
\omega_P = e^{-u}\big( dt -\im d\xi_{jb}\xi^{jb}  \big) = \triangle \tau \;\; \Rightarrow \;\; D^{ia}t - D^{ia}\xi_{jb}\xi^{jb} =0, \;\; \dot t - \im \dot\xi_{ia}\xi^{ia} = e^u, \nn \\
\big(\omega_Q\big)_{ia} = e^{-u/2}\big( e^v \big)_i^j \big( e^\varphi \big)_a^b \big( d\xi_{jb} - \triangle t \psi_{jb}  \big) = d\theta_{ia}\;\; \Rightarrow \;\; D^{jb}\xi_{ia} = e^{u/2}\big( e^{-v} \big)_i^j \big( e^{-\varphi} \big)_a^b, \;\; \psi_{ia} = e^{-u} \dot\xi_{ia}.
\eea
As a result of \p{maincondcons}, $D^{ia}\xi_{jb}$ satisfies relations
\be\label{Dxirels1}
D_{jb}\xi^{ia} = e^{u/2} \big( e^{v} \big)_j^i \big( e^{\varphi} \big)_b^a \;\; \Rightarrow \;\; D^{ia}\xi_{kc} D_{jb}\xi^{kc} = \delta^i_j \delta^a_b e^u, \;\; D^{kc}\xi_{jb} D_{kc}\xi^{ia} = \delta^i_j \delta^a_b e^u,
\ee
and, moreover,
\be\label{Dxirels2}
D_{kc}\xi^{ia}\, D^{kb}\xi_{ld} = \frac{1}{2}\delta_l^i \, D_{kc}\xi^{ma}D^{kb}\xi_{md}, \;\; \mbox{etc.}
\ee
Using these relations, it is possible to find $D_{ld}e^u$
\be\label{expu}
e^u = \frac{1}{4} D^{ia}\xi_{kc} D_{ia}\xi^{kc} \;\; \Rightarrow \;\; D_{ld}e^u = 2\im \dot\xi{}_{ia}D_{ld}\xi^{ia}.
\ee

The super--Schwarzian $\sigma_{ia}$ can be obtained as a $d\theta$ -projection of either the forms $\omega_T$ or $\omega_J$. For example, $T$ part of the Cartan form reads
\be\label{omegaTschw}
\im \big( \omega_T \big)_{ij} T^{ij} = \im \triangle\tau \big( \cS_T  \big)_{ij} T^{ij} -\frac{2\im}{3} T^{ij}d\theta_{ia}\sigma^a_j  =  -\im T_k{}^m \left( e^{-v}\right)_i^k d\left(e^v\right)_m^i + \im T^{km} \big( e^v  \big)_k^i \big( e^v  \big)_m^j \left(\hat\omega_T\right)_{ij},
\ee
where $\omega_{ij}$ is given by \p{su2CAF}.

To obtain $d\theta$ -projection of $\left( e^{-v}\right)_i^k d\left(e^v\right)_m^i$, one should note that due to \p{maincondcons},
\be\label{DDxiDxi}
D_{ld}D_{kc}\xi^{ia} D^{jb}\xi_{ia} = \frac{1}{2}\delta_k^j \delta_c^b D_{ld}e^u + e^u \delta_c^b \big(  e^{-v} \big)_i^j D_{ld} \big(e^{v}\big)_k^i + e^u \delta_j^k \big( e^{-\varphi}  \big)_a^b D_{ld} \big( e^\varphi \big)_c^a.
\ee
Substituting this into relation
\be\label{DDxiDxi2}
D_{ld}D_{kc}\xi^{ia} D^{jb}\xi_{ia} = \big\{ D_{kc},D_{ld}  \big\}\xi^{ia} D^{jb}\xi_{ia} - D_{kc}D_{ld}\xi^{ia} D^{jb}\xi_{ia}
\ee
and taking trace over $c$, $b$, one can find
\bea\label{DDxiDx3}
2 e^u \big( e^{-v}  \big)_i^j D_{ld} \big( e^v \big)_k^i + e^u\big( e^{-v}  \big)_i^j D_{kd} \big( e^v \big)_l^i &=& -2\im \delta^j_k \dot\xi_{ia}D_{ld}\xi^{ia} -\im \delta^j_l \dot\xi_{ia}D_{kd}\xi^{ia} -\nn \\&&-2\im \epsilon_{kl} \dot\xi_{ia}D^j_{d}\xi^{ia}- \delta_l^j e^u \big( e^{-\varphi}\big)_a^b D_{kb}\big( e^\varphi \big)_d^a.
\eea
Therefore,
\bea\label{DDxiDx4}
e^u \big( e^{-\varphi} \big)_a^b D_{kb} \big( e^\varphi  \big)_d^a = - e^u \big(  e^{-v}  \big)_i^j D_{jd} \big( e^v  \big)_k^i -3\im \dot\xi_{ia} D_{kd}\xi^{ia}, \nn \\
e^u \big(  e^{-v}  \big)_i^j D_{ld} \big( e^v  \big)_k^i  = \frac{2}{3} \delta_l^j e^u \big(  e^{-v}  \big)_n^m D_{md} \big( e^v  \big)_k^n - \frac{1}{3} \delta_k^j e^u \big(  e^{-v}  \big)_n^m D_{md} \big( e^v  \big)_l^n, \nn \\
e^u \big(  e^{-v}  \big)_n^m D_{md} \big( e^v  \big)_k^n = \frac{1}{2}D_{jd}D_{kb}\xi^{ia}\, D^{jb}\xi_{ia} - \im \dot\xi^{ia}D_{kd}\xi_{ia}.
\eea
The rest of the form reads
\bea\label{omegaTrest}
&&-d\theta^{ld}(1+\alpha)T^{km}\big( e^v  \big)_k^i \big( e^v \big)_m^j \left(  \psi_{ia} D_{ld} \xi^a_j + \psi_{ja} D_{ld} \xi^a_i  \right) =\nn \\
&&= 2(1+\alpha)T^{km} d\theta_{kd}  \dot\xi^a_i \big(  e^{-\varphi} \big)_a^d \big( e^v \big)_k^i e^{-u/2} = 2(1+\alpha)T^{km}d\theta_{kd} e^{-u} {\dot\xi}^{ia}D^d_m \xi_{ia}.
\eea
These results ensure that $d\theta$-projection of the form $\omega_T$ has the structure \p{omegaJT}, with $\sigma_{kd}$ being
\bea\label{sigmaT}
\sigma_{dk} = \frac{1}{4} e^{-u} \left[ D_{jd}, D_{ck}\right] \xi^{ia} D^{jc}\xi_{ia} +\frac{3}{2} \im \left(1+ 2\alpha\right) e^{-u} {\dot\xi}^{ia} D_{dk}\xi_{ia}.
\eea
Analogous study of form $\omega_J$
\be\label{omegaTschw}
\im \big( \omega_J \big)_{ab} J^{ab} = \im \triangle\tau \big( \cS_J  \big)_{ab} J^{ab} +\frac{2\im}{3} J^{ab}d\theta_{ia}\sigma^i_b  =  -\im J_c{}^d \left( e^{-\varphi}\right)_a^c d\left(e^\varphi\right)_d^a + \im J^{cd} \big( e^{\varphi}  \big)_c^a \big( e^\varphi  \big)_d^b {\hat\omega}_{ab},
\ee
leads to the same expression \p{sigmaT}.

Thus, we see  that all the Cartan forms expressed through the fermionic superfield $\sigma_{ai}$ \p{sigmaT}.
We associate this field with $\cN{=}4$ super--Schwarzian we are looking for:
\be
\cS(\tau, \theta)_{ia} = \frac{\left[ D_{ja}, D_{ci}\right] \xi^{kb} D^{jc}\xi_{kb}}{D^{md} \xi_{ne} D_{md}\xi^{ne}}+
6 \im \left(1+ 2\alpha\right) \frac{ {\dot\xi}^{dk} D_{ia}\xi_{dk}}{D^{md} \xi_{ne} D_{md}\xi^{ne}}.
\ee

\section{$\cN=4$ Schwarzian action}
Like the previously considered cases \cite{KK}, one may ask whether the superfield Schwarzian action, which provides the $D(1,2;\alpha)$ - invariant generalization of the bosonic Schwarzian action,
\be\label{bosact}
S = - \frac{1}{2} \int d\tau \cS(t,\tau) = - \frac{1}{2}\int d\tau \left(  \frac{\dddot t}{\dot t} - \frac{3}{2} \left( \frac{\ddot t}{\dot t}\right)^2  \right),
\ee
could be constructed. As is shown in the Section $3$ and the Appendix C, the Maurer--Cartan equations imply that the only superfields, invariant with respect to $D(1,2;\alpha)$ group transformations, are the super--Schwarzian $\sigma^{ia}$ and its derivatives. Therefore, it would be natural to expect that the superfield action is some integral of $\sigma^{ia}$ over the part of superspace. Indeed, let us show that the expression
\be\label{N4actcand}
S = -\frac{1}{72}\int d\tau D^{kc}D_{lc}D_{kb}\sigma^{lb} |_{\theta \rightarrow 0}
\ee
is invariant with respect to $\cN=4$ supersymmetry, realized on superspace coordinates $\tau$ and $\theta_{ia}$ as
\be\label{tauthetavar}
\delta \tau = -\im \epsilon^{ia}\theta_{ia}, \;\; \delta\theta_{ia} = \epsilon_{ia}, \;\; \delta\triangle\tau=0, \;\; \delta d\theta_{ia} =0.
\ee
The active variation of any superfield $f$ with respect to transformations \p{tauthetavar} is given by the formula
\be\label{sfvar}
\delta^\star f = -\delta \tau \frac{\partial f}{\partial \tau}  -\delta\theta_{ia} \frac{\partial f}{\partial \theta_{ia}} \equiv - \epsilon_{ia} {\widehat Q}{}^{ia} f,\;\; {\widehat Q}{}^{ia} = \frac{\partial}{\partial \theta_{ia}} - \im \theta^{ia} \frac{\partial}{\partial \tau}.
\ee
It can be straightforwardly shown that the differential operator ${\widehat Q}{}^{ia}$ anticommutes, as expected, with the covariant derivative $D^{jb}$ and differs from it by the sign of the $\theta\partial_\tau$ -term. Therefore,
\be\label{N4actcandvar1}
\delta^\star S = \frac{1}{72}\int d\tau D^{kc}D_{lc}D_{kb}\epsilon_{ia}{\widehat Q}{}^{ia} \sigma^{lb} |_{\theta \rightarrow 0} = \frac{1}{72}\epsilon_{ia}\int d\tau {\widehat Q}{}^{ia} D^{kc}D_{lc}D_{kb} \sigma^{lb} |_{\theta \rightarrow 0}.
\ee
As after applying differential operators on $\sigma^{ia}$ we take limit $\theta \rightarrow 0$, the $\theta\partial_\tau$ -term in ${\widehat Q}{}^{ia}$ is irrelevant, and ${\widehat Q}{}^{ia}$ can be replaced with $D^{ia}$. Therefore
\be\label{N4actcandvar2}
\delta^\star S  = \frac{1}{72}\epsilon_{ia}\int d\tau D^{ia} D^{kc}D_{lc}D_{kb} \sigma^{lb} |_{\theta \rightarrow 0}= \frac{1}{72}\epsilon_{ia}\int d\tau \Big[ 2\im D^{ia}D^{kc}{\dot\sigma}_{kc} - \im D^a_l \big( D^{ib}{\dot\sigma}_b^l + D^{lb}{\dot\sigma}_b^i \big)   \Big]|_{\theta \rightarrow 0}=0,
\ee
where the expression for $D^{ia} D^{kc}D_{lc}D_{kb} \sigma^{lb}$ is a consequence of the constraint \p{sigmaconstr}.

The supersymmetry invariant integral over $d\tau$ can be presented as an integral over part of the superspace:
\be\label{N4actfin}
S = \frac{1}{72}\int d\tau d\theta^{kc}d\theta_{lc}d\theta_{kb}\sigma^{lb}.
\ee

One can also evaluate the component form  of this action. Simplest way to do so is to observe that the $\triangle \tau$ projection of the form $\omega_K = \triangle\tau C + \ldots$ \p{formssol} contains third derivative of $\sigma^{ia}$ (this expression can be found in the Appendix C \p{C}). Comparing this with the projection which can be obtained directly from \p{scalarCF} after applying all the necessary conditions, one can obtain that
\bea\label{N4actcomp}
S = -\frac{1}{72}\int d\tau D^{kc}D_{lc}D_{kb}\sigma^{lb} |_{\theta \rightarrow 0} = -\int d\tau \left[  \frac{1}{2}\alpha (1+\alpha) \left( \frac{\partial_\tau \big( \ddot t - \im \ddot\xi_{ia}\xi^{ia}  \big)}{\dot t - \im \dot\xi_{jb}\xi^{jb}}   - \frac{3}{2} \frac{\big( \ddot t - \im \ddot\xi_{ia}\xi^{ia}  \big)^2 }{ \big( \dot t - \im \dot\xi_{jb}\xi^{jb} \big)^2}\right) \right. + \nn \\+\left. \im \alpha(1+\alpha) \frac{\dot\xi_{ia}{\ddot\xi}{}^{ia}}{\dot t - \im \dot\xi_{jb}\xi^{jb}} - \frac{1}{2}\alpha  \partial_\tau  \big( e^{-v}   \big)_l^i \partial_\tau  \big( e^{v}   \big)_i^l  + \frac{1}{2}(1+\alpha)  \partial_\tau \big( e^{-\phi}   \big)_d^a  \partial_\tau \big( e^{\phi}   \big)_a^d \right. - \nn \\ \left. -\im \alpha(1+\alpha) \frac{\big( e^{-v}  \big)_k^j \partial_\tau \big( e^v  \big)_j^l \dot\xi_{lb}\dot\xi{}^{kb}}{\dot t - \im \dot\xi_{ia}\xi^{ia} } -\im\alpha(1+\alpha) \frac{\big( e^{-\phi}  \big)_c^b \partial_\tau \big( e^\phi  \big)_b^d \dot\xi_{jd}\dot\xi{}^{jc}}{\dot t - \im \dot\xi_{ia}\xi^{ia} } - \frac{\im}{9} \dot\sigma{}^{kc}\sigma_{kc} - \frac{1}{9}F^2 \right].
\eea
Here, we denote the first component of each superfield with the same letter, and $4F = D^{mc}\sigma_{mc}|_{\theta\rightarrow 0}$. Note that the first component of the super--Schwarzian $\sigma^{ia}$ can be treated as an independent one, as $D_{ia}D_{jb}\xi^{kc}$ can not be expressed in terms of time derivatives of anything else. The same applies to $D^{mc}\sigma_{mc}$, too.

The action \p{N4actcomp} is invariant with respect to the whole $D(1,2;\alpha)$ group for general $\alpha$ and should contain $SU(1,1|2)$ case, which corresponds to $\alpha =0$ or $-1$, as a particular limit. However, simply setting $\alpha =0$ or $-1$ in \p{N4actcomp} would remove the most important terms in the action. To take the limit properly, we should, at first, ``renormalize'' the action by dividing it by $\alpha(1+\alpha)$, thus removing $\alpha$ dependence from the most of the terms. Secondly, one should remove $\sigma_{kc}|_{\theta\rightarrow 0}$ and $F$ by their equations of motion, $\dot\sigma{}^{ia}|_{\theta\rightarrow 0} =0$ and $F=0$. Thirdly, one should set to zero $\phi^{ab}$, if limit $\alpha\rightarrow 0$ is to be taken, or $v^{ij}$ if $\alpha\rightarrow -1$. Then the action becomes nonsingular in $\alpha$ and after taking the appropriate limit coincides with one obtained in \cite{KK}. Note that somewhat confusing difference in signs of kinetic terms of $v^{ij}$ and $\phi^{ab}$ allows to obtain proper sign of the kinetic term of the remaining field in the $SU(1,1|2)$ action for $\alpha=0$ and $\alpha=-1$.

\setcounter{equation}0
\section{Conclusion}
In this work we applied  the method of nonlinear realizations to the construction of the $\cN{=}4$ super--Schwarzian
associated with the $D(1,2;\alpha)$ conformal group. As compared to the previous attempt to utilize the nonlinear realizations for construction of the $\cN{=}4$ super--Schwarzians \cite{Gal5} we successfully used the minimal set
of the constraints on the Cartan forms advocated in \cite{KK}:
\begin{itemize}
	\item For the superalgebra containing the super Poincar\'{e} subalgebra $\left\{ Q{}^{ia}, Q{}^{jb} \right\} = -2\epsilon^{ij}\epsilon^{ab} P$ the invariant super-space $\left\{\tau, \theta_{ia}\right\}$ defined as
	$$ \omega_P = d\tau - \im d\theta^{ia} \theta_{ia}, \; \big(\omega_Q\big){}_{ia} = d \theta{}_{ia} \qquad \qquad (a)$$
	\item The unique additional constraint has to be imposed on the Cartan form for dilatation
	$$\omega_D=0 \qquad \qquad (b).$$
\end{itemize}	
From the general structure of the Cartan forms upon imposing the constraints (a,b), it follows that the fermionic components of the forms in \p{eq1}, \p{eq2}, \p{eq3} and  \p{eq4} are quite nontrivial.
Therefore, any constraint would be imposed on these forms will result in the constraints on the super--Schwarzian $\sigma_{ia}$. That is why our minimal set of the constraints is the maximally possible one.
We also demonstrated that the Maurer--Cartan equations greatly simplified all calculations helping to express all Cartan forms in terms of the single object --- $\cN{=}4$ super--Schwarzian. However, to find the expression of the  $\cN{=}4$ super--Schwarzian in terms of the basic superfields one has to again use all set of constraints.

We are planning to apply the proposed approach to $\cN$-extended superconformal group including the variant of $OSp(n|2)$ superconformal symmetry. Another  interesting problem is to obtain non-relativistic and/or Carrollian versions of the Schwarzian \cite{gomis}, as well as to the flat space analogue of the Schwarzian \cite{FlatSch}.

\section*{Acknowledgements}

The work was supported by Russian Foundation for Basic Research, grant
No~20-52-12003.
\setcounter{equation}0
\def\theequation{A.\arabic{equation}}
\section*{Appendix A. Superalgebra $D(1,2;\alpha)$}
The set of the generators spanning $D(1,2;\alpha)$ superalgebra includes
\bea
\mbox{ Bosonic generators: } && P,D,K - \mbox{ forming } sl(2) \mbox{ algebra} \nn \\
&& \mbox{the $su(2) \times su(2)$ generators} \; T^{ij}=T^{ji}, J^{ab}=J^{ba}, \; i,j =1,2; \; a,b = 1,2  \nn\\
\mbox{ Fermionic generators: } && Q^{ia}, S^{ia} ,
\eea
which obey the following conjugation rules
\be
\left(T^{ij}\right)^\dagger = T_{ij}, \; \left(J^{ab}\right)^\dagger = J_{ab},\; \left(P,D,K\right)^\dagger = (P,D,K),  \qquad \left(Q^{ia}\right)^\dagger = Q_{ia}, \; \left(S^{ia}\right)^\dagger = S_{ia}.
\ee
The non-zero (anti)commutators are
\bea\label{D12}
&& \im \left[ P,K\right] = - 2 D,  \; \im \left[ P,D\right] =-P, \;
\im \left[K,D\right]=K,  \nn \\
&& \im  \left[ T^{ij}, T^{km}\right] = \epsilon^{ik} T^{jm}+ \epsilon^{jm} T^{ik}, \quad \im \left[ J^{ab},J^{cd}\right] = \epsilon^{ac} J^{bd}+\epsilon^{bd}J^{ac}, \nn \\
&& \im \left[ P, S^{ia}\right] =- Q^{ia}, \;  \im \left[ K , Q^{ia}\right] = S^{ia},\quad  \im \left[ D , Q^{ia}\right] = \frac{1}{2} Q^{ia}, \;
\im \left[ D , S^{ia}\right] = -\frac{1}{2} S^{ia}, \nn \\
&& \im  \left[ T^{ij}, Q^{ka}\right] = \frac{1}{2} \left( \epsilon^{ik} Q^{ja}+ \epsilon^{jk} Q^{ia}\right), \;
\im  \left[ J^{ab}, Q^{ic}\right] = \frac{1}{2} \left( \epsilon^{ac} Q^{ib}+ \epsilon^{bc} Q^{ia}\right), \nn \\
&& \im  \left[ T^{ij}, S^{ka}\right] = \frac{1}{2} \left( \epsilon^{ik} S^{ja}+ \epsilon^{jk} S^{ia}\right), \;
\im  \left[ J^{ab}, S^{ic}\right] = \frac{1}{2} \left( \epsilon^{ac} S^{ib}+ \epsilon^{bc} S^{ia}\right), \nn \\
&& \left\{ Q^{ia}, Q^{jb}\right\} = -2 \epsilon^{ij} \epsilon^{ab} P, \;
\left\{ S^{ia}, S^{jb}\right\} = -2 \epsilon^{ij} \epsilon^{ab} K, \nn \\
&&\left\{ Q^{ia}, S^{jb}\right\} = 2 \left( -\epsilon^{ij} \epsilon^{ab} D+
\alpha \epsilon^{ij} J^{ab} - (1+\alpha) \epsilon^{ab} T^{ij}\right).
\eea

\setcounter{equation}0
\def\theequation{B.\arabic{equation}}
\section*{Appendix B. $su(2)$ rotations}
Using the commutator relations of the $D(1,2;\alpha)$ algebra \p{D12} it is not too complicated to find the effect
of the  $su(2)\times su(2)$ rotations on the fermionic and $su(2)$ generators
\bea
e^{-\im v \cdot T}\, Q^{kc}\, e^{\im v \cdot T}& = & \left( e^{v}\right)^k_m Q^{mc} =
\cos \sqrt{\frac{v^2}{2}}Q^k+\frac{\sin\sqrt{\frac{v^2}{2}}}{\sqrt{\frac{v^2}{2}}} v^k_m Q^{mc}, \qquad v^2 \equiv v_{ij}v^{ij}, \label{Q1}\\
e^{-\im \phi \cdot J}\, Q^{kc}\,e^{\im \phi \cdot J} & = & \left( e^{\phi}\right)^c_d Q^{kd} =
\cos \sqrt{\frac{\phi^2}{2}}Q^{kc}+\frac{\sin\sqrt{\frac{\phi^2}{2}}}{\sqrt{\frac{\phi^2}{2}}} \phi^c{}_d Q^{kd}, \qquad \phi^2 \equiv \phi_{ab}\phi^{ab}, \label{Q2}\\
e^{-\im v\cdot T}\, T^{km}\,e^{\im v\cdot T} & = & T^{km} +\frac{\sin\sqrt{2 v^2}}{\sqrt{2 v^2}}\left( v^k_n T^{nm}+v_n^m T^{nk} \right)+\frac{1-\cos\sqrt{2 v^2}}{2 v^2}\left( - v^2 T^{km}+2 v^k_i v^m_j T^{ij}\right), \label{T1}\\
e^{-\im \phi\cdot J}\, J^{cd}\,e^{\im \phi\cdot J} & = & J^{cd} +\frac{\sin\sqrt{2 \phi^2}}{\sqrt{2 \phi^2}}\left( \phi^c_b J^{bd}+\phi^d_b J^{bc} \right)+\frac{1-\cos\sqrt{2 \phi^2}}{2 \phi^2}\left( - \phi^2 J^{cd}+2 v^c_a v^d_b J^{ab}\right), \label{T2}
\eea
and
\bea
e^{-\im v\cdot T} d e^{\im v \cdot T} & = &
\im d v_{km} \left[T^{km}+\frac{1-\cos\sqrt{2 v^2}}{ v^2} v^k_i T^{im}+\frac{\sqrt{2 v^2}-\sin\sqrt{2 v^2}}{2 v^2 \sqrt{ 2 v^2}}\left(- v^2 T^{km}+2 v_i^k v_j^m T^{ij}\right)\right],\label{dT1} \\
e^{-\im \phi\cdot J} d e^{\im \phi \cdot J} & = &
\im d \phi_{ab} \left[J^{ab}+\frac{1-\cos\sqrt{2 \phi^2}}{ \phi^2} \phi^a_d J^{db}+\frac{\sqrt{2 \phi^2}-\sin\sqrt{2 \phi^2}}{2 \phi^2 \sqrt{ 2 \phi^2}}\left(- \phi^2 J^{ab}+2 \phi_c^a \phi_d^b J^{cd}\right)\right]. \label{dT2}
\eea
It is less evident to note that the expressions \p{T1} and \p{T2}  can be written, similarly to \p{Q1} and \p{Q2}, as
\be
e^{-\im v\cdot T} T^{km} e^{\im v \cdot T} = \left( e^v\right)^k_i T^{ij}\left(e^v\right)_j^m \quad
\mbox{and} \quad
e^{-\im \phi\cdot J} J^{cd} e^{\im \phi \cdot J} = \left( e^\phi\right)^c_a J^{ab}\left(e^\phi\right)_b^d.
\ee
Finally, the expressions \p{dT1} and \p{dT2} can be also written in a simplified way as
\be
e^{-\im v\cdot T} d e^{\im v \cdot T} = -\im \left(T\right){}_k{}^m \left( e^{-v}\right)_i^k d\left(e^v\right)_m^i
\quad \mbox{and} \quad
e^{-\im \phi\cdot J} d e^{\im \phi \cdot J} = -\im \left(J\right){}_c{}^b \left( e^{-\phi}\right)_a^c d\left(e^\phi\right)_b^a .
\ee

Note the useful identities which simplify the explicit calculations
\bea
\left(e^v\right)^k_m =\cos\sqrt{\frac{v^2}{2}}\delta^k_m+\frac{\sin\sqrt{\frac{v^2}{2}}}{\sqrt{\frac{v^2}{2}}} v^k_m ,\quad   \left(e^{-v}\right)^k_m =\cos\sqrt{\frac{v^2}{2}}\delta^k_m-\frac{\sin\sqrt{\frac{v^2}{2}}}{\sqrt{\frac{v^2}{2}}} v^k_m = -\epsilon^{ki}\epsilon_{mj}  \left(e^v\right)^j_i.
\eea

\setcounter{equation}0
\def\theequation{C.\arabic{equation}}
\section*{Appendix C. Solution to the Maurer--Cartan equations}
As we already demonstrated the  equation $d\omega_P$ \p{MCexp} is satisfied due to the condition \p{maincond}.
In contrast, $d\omega_Q$ equation is not trivial. After substitution of \p{maincond}, it separates into two equations:
\bea\label{domegaQ}
\big( \triangle_1 \tau d_2 \theta_{jb} - \triangle_2 \tau d_1 \theta_{jb}   \big):&& 0= -A_{ia}{}^{jb} - \delta_i^j \big( \cS_J\big)_a{}^b - \delta_a^b \big( \cS_T\big)_i{}^j, \nn \\
d_1\theta_{kc}d_2\theta_{ld}: &&0= \delta_i^l \Sigma_a{}^{d|kc} + \delta_i^k \Sigma_a{}^{c|ld} + \delta_a^d \Pi_i^{l|kc} + \delta_a^c \Pi_i{}^{k|ld}.
\eea
The first of these equations straightforwardly expresses $A_{ia}{}^{jb}$ in terms of $\big( \cS_J\big)_a{}^b$ and $\big( \cS_T\big)_i{}^j$, the second one is more elaborate. At first, multiplying it by $\delta_l^i \delta_c^a$, one can obtain
\be\label{SigmaPi1}
\Sigma_c{}^{d|kc} + \Pi_l{}^{k|ld} =0 \;\; \Rightarrow \;\; \Sigma_c{}^{d|kc} = \sigma^{kd}, \;\;  \Pi_l{}^{k|ld} =- \sigma^{kd}.
\ee
Next, multiplying by just $\delta_l^i$ and taking into account \p{SigmaPi1}, one obtains
\be\label{SigmaPi2}
2\Sigma_a{}^{d|kc} + \Sigma_{a}{}^{c|kd} - \delta_a^c \sigma^{kd} =0 \;\; \Rightarrow \;\; \Sigma_a{}^{d|kc} = \frac{2}{3}\delta_a^c \sigma^{kd} - \frac{1}{3}\delta_a^d \sigma^{kc}.
\ee
Multiplying by $\delta_d^a$, one obtains
\be\label{SigmaPi3}
2\Pi_i{}^{l|kc} + \Pi_i{}^{k|lc} + \delta_i^k \sigma^{lc}=0\;\; \Rightarrow  \Pi_i{}^{k|lc} = -\frac{2}{3} \delta_i^l \sigma^{kc} + \frac{1}{3}\delta_i^k \sigma^{lc}.
\ee
Substituting these relations back into \p{domegaQ}, one notes that $d\theta\times d\theta$ equation is satisfied with no further constraints on $\sigma^{kd}$, and $\omega_J$ and $\omega_T$ forms can be written as
\be\label{omegaJT}
\big(\omega_J \big)_{ab} = \triangle\tau \big( \cS_J   \big)_{ab} + \frac{1}{3} \big( d\theta_{ia}\sigma^i_b + d\theta_{ib}\sigma^i_a  \big), \;\; \big(\omega_T \big)_{ij} = \triangle\tau \big( \cS_T   \big)_{ij} - \frac{1}{3} \big( d\theta_{ia}\sigma^a_j + d\theta_{ja}\sigma^a_i  \big).
\ee
The fermion $\sigma_{ia}$ is an obvious candidate for the super--Schwarzian. Note that if $\alpha =0,-1$, the generators of one $SU(2)$ groups do not appear at the right hand side of commutators of supercharges. If this decoupled $SU(2)$ is dropped entirely from the coset space, equation \p{domegaQ} would not contain either $\Sigma$ or $\Pi$. This equation would, as it follows from \p{SigmaPi1}, set the remaining fermion to zero and, as was already found, the bosonic component of the automorphism form becomes the super--Schwarzian. We, therefore, assume that $\alpha\neq 0,-1$ in further considerations.

Not all the equations have been written down. The $d\omega_J$ equation also separates into two:
\bea\label{domegaJ}
\triangle_1 \tau d_2\theta_{kc} - \triangle_2\tau d_1\theta_{kc} : && \im D^{kc}\big( \cS_J  \big)_{ab} - \frac{\im}{3} \big( \delta_a^c {\dot\sigma}_b^{k} + \delta_b^c {\dot\sigma}_a^{k}   \big) = -\frac{\im}{3} \left(   \delta_b^c \big( \cS_J  \big)_{a}{}^d \sigma^k_d + \delta_a^c \big( \cS_J  \big)_{b}{}^d \sigma^k_d \right)-\nn \\&& - \frac{\im}{3} \left( \big( \cS_J  \big)_{a}{}^c \sigma^k_b + \big( \cS_J  \big)_{b}{}^c \sigma^k_a  \right) + \alpha \left( \delta_b^c \Psi_a^k + \delta_a^c \Psi_b^k  \right), \\
d_1\theta_{kc} d_2\theta_{ld}: && 2\epsilon^{kl}\epsilon^{cd} \big( \cS_J \big)_{ab} + \frac{\im}{3} \left( \delta_a^c D^{ld}\sigma_b^k + \delta_b^c D^{ld}\sigma_a^k + \delta_a^d D^{kc}\sigma_b^l + \delta_b^d D^{kc}\sigma^l_a \right) = \nn \\
&& =\epsilon^{kl}\epsilon^{cd} \left( -\frac{2\im}{9} \sigma_{ma}\sigma^m_b  -2\alpha \big( \cS_J  \big)_{ab}  \right) + \big( \delta_a^c \delta_b^d + \delta_b^c \delta_a^d   \big) \left(2\alpha \big( \cS_T \big)^{kl} - \frac{2\im}{9} \sigma^k_f \sigma^{lf} \right). \nn
\eea
Substituting $D^{ia}\sigma^{jb}$ into the second equation as most general combination of tensors of various symmetries
\bea\label{DsigmaF}
D^{ia}\sigma^{jb} = \epsilon^{ij}\epsilon^{ab}F - \frac{1}{2}\epsilon^{ij}F^{(ab)}- \frac{1}{2}\epsilon^{ab}F^{(ij)} + F^{(ij)(ab)}, \nn \\
F = \frac{1}{4}D^{kc}\sigma_{kc}, \;\; F^{ij} = \frac{1}{2}\big( D^{ic}\sigma_c^j +D^{jc}\sigma_c^i  \big), \;\; F^{ab} = \frac{1}{2}\big( D^{ka}\sigma_k^b +D^{kb}\sigma_k^a  \big),
\eea
one can obtain that second equation \p{domegaJ} implies $F^{(ij)(ab)} =0 $, relates $F^{ij}$ and $F^{ab}$ to $\big( \cS_T\big)^{ij}$ and $\big( \cS_J  \big)^{ab}$
\be\label{SJST}
\big(\cS_J \big)^{ab} = \frac{\im}{1+\alpha} \left[ \frac{F^{ab}}{6} - \frac{\sigma^a_m \sigma^{mb}}{9}   \right], \;\; \big(\cS_T \big)^{ij} = \frac{\im}{\alpha} \left[ \frac{F^{ij}}{6} + \frac{\sigma^i_c \sigma^{jc}}{9}   \right],
\ee
and places no restriction on the scalar $F$.
Therefore, $\sigma^{ia}$ satisfies the differential constraint
\be\label{Dsigmarel}
D^{ia}\sigma^{jb} +D^{ja}\sigma^{ib} +D^{ib}\sigma^{ja} +D^{jb}\sigma^{ia} =0, \quad
D^{ia} \sigma^{jb}+D^{jb}\sigma^{ia} = \frac{1}{2} \epsilon^{ij} \epsilon^{ab} D^{kc} \sigma_{kc} .
\ee
To study the first equation of \p{domegaJ}, one should find the derivative of $\big(\cS_J \big)^{ab}$, and, therefore, of $F$, $F^{ij}$ and $F^{ab}$. Using their definition \p{DsigmaF} and commutation relation of derivatives \p{covders}, one can obtain the relations
\bea\label{DFFF}
D^{kc}F^{ab} = -\frac{1}{3}\epsilon^{ac}D^k_d F^{bd} -\frac{1}{3}\epsilon^{bc}D^k_d F^{ad}, \;\; D^{kc}F^{ij} = -\frac{1}{3}\epsilon^{ik}D^c_l F^{jl} -\frac{1}{3}\epsilon^{jk}D^c_l F^{il}, \nn \\
D^{kc}F = \im {\dot\sigma}{}^{kc}, \;\; D^i_c F^{ac} = 6\im {\dot\sigma}{}^{ia} - D^a_k F^{ik}.
\eea
We prefer to express derivatives of $F^{cd}$ in terms of derivatives of $F^{ij}$.
Substituting $\big(\cS_J   \big)^{ab}$ \p{SJST} into first equation \p{domegaJ} and evaluating derivatives, it could be obtained that all the terms neither proportional to $\delta_a^c$ or $\delta_b^c$ vanish and the rest imply that
\be\label{Psi}
\alpha(1+\alpha)\Psi^{ia} = - \frac{\im}{3} \alpha {\dot\sigma}{}^{ia} - \frac{1}{18}D^a_j F^{ij} + \frac{1}{9}F \sigma^{ia} - \frac{1}{18} F^{il}\sigma_l^a - \frac{1}{18} F^{ac}\sigma_c^i + \frac{4}{81} \sigma^i_c \sigma_m^c \sigma^{ma}.
\ee

Equation $d\omega_T$ is very much similar to $d\omega_J$:
\bea\label{domegaT}
\triangle_1 \tau d_2\theta_{kc} - \triangle_2\tau d_1\theta_{kc} : && \im D^{kc}\big( \cS_T  \big)_{ij} + \frac{\im}{3} \big( \delta_i^k {\dot\sigma}_j^{c} + \delta_j^k {\dot\sigma}_i^{c}   \big) =\frac{\im}{3} \left(   \delta_j^k \big( \cS_T  \big)_{i}{}^l \sigma^c_l + \delta_i^k \big( \cS_T  \big)_{j}{}^l \sigma^c_l \right)-\nn \\&& + \frac{\im}{3} \left( \big( \cS_T  \big)_{i}{}^k \sigma^c_j + \big( \cS_T  \big)_{j}{}^k \sigma^c_i  \right) -(1+ \alpha) \left( \delta_i^k \Psi_j^c + \delta_j^k \Psi_i^c  \right), \\
d_1\theta_{kc} d_2\theta_{ld}: && 2\epsilon^{kl}\epsilon^{cd} \big( \cS_T \big)_{ij} - \frac{\im}{3} \left( \delta_i^k D^{ld}\sigma_j^c + \delta_j^k D^{ld}\sigma_i^c + \delta_i^l D^{kc}\sigma_j^d + \delta_j^l D^{kc}\sigma_i^d \right) = \nn \\
&& = \epsilon^{kl}\epsilon^{cd} \left( -\frac{2\im}{9} \sigma_{ia}\sigma^a_j  + 2(1+\alpha) \big( \cS_T  \big)_{ij}  \right) + \big( \delta_i^k \delta_j^l + \delta_j^k \delta_i^l  \big) \left( -2(1+ \alpha) \big( \cS_J \big)^{cd} - \frac{2\im}{9} \sigma^c_m \sigma^{md} \right). \nn
\eea
Substitution of relations obtained above \p{DsigmaF}, \p{Dsigmarel}, \p{SJST}, \p{Psi} guarantees that these two equations are satisfied in the same manner as $d\omega_J$ equations \p{domegaJ}.

The equation $d\omega_D$ results in the following relations:
\be\label{omegaD}
\Gamma^{ia} = \im \Psi^{ia}, \quad A^{ia|jb}+ A^{jb|ia}=0.
\ee
The second equation in \p{omegaD} is satisfied after substitution $A^{ia|jb} = -\epsilon^{ij} \big( \cS_J \big)^{ab} - \epsilon^{ab}\big( \cS_T  \big)^{ij}$ \p{domegaQ}, while the first one expresses the
$\Gamma^{ia}$ through $\Psi^{ia}$.

Now, combining everything together we will have the following expressions for the Cartan forms
\bea
&& \omega_P = \triangle \tau, \quad \omega_D=0,\quad \omega_K = \triangle \tau C + \im d\theta_{ia} \Psi^{ia}, \\
&& \left(\omega_J\right)_{ab}  = \triangle \tau \left(\cS_J\right)_{ab}+\frac{1}{3}\left( d\theta_{ka} \sigma^k_b+
d\theta_{kb} \sigma^k_a \right), \\
&& \left(\omega_T\right)_{ij}  = \triangle \tau \left(\cS_T\right)_{ij}-\frac{1}{3}\left( d\theta_{ic} \sigma^c_j+
d\theta_{jc} \sigma^c_i \right),\\
&& \left(\omega_Q\right)^{ia} = d \theta^{ia}, \quad
\left(\omega_S\right)^{ia} = \triangle \tau \Psi^{ia} -d\theta_{ib}  \left(\cS_J\right)^b_{a}
-d\theta_{ka}  \left(\cS_T\right)^k_{i} ,
\eea
with only one function $C$ remaining to be determined by study of $d\omega_S$ and $d\omega_K$ equations.

The $d\omega_S$ equation again separates into two
\bea
D^{jb} \Psi^{ia}+\epsilon^{ij} \left( \dot{\cS}_J\right){}^{ab} + \epsilon^{ab} \left( \dot{\cS}_T\right){}^{ij} =
-\frac{1}{2} \epsilon^{ij}\epsilon^{ab} \left( \cS_J^2 + \cS_T^2 - 2 C\right) + 2 \left( \cS_J\right)^{ab} \left( \cS_T\right)^{ij} -\nn \\ -\frac{1}{3}\left( \epsilon^{ab} \big( \sigma^j_c \Psi^{ic} + \sigma^i_c \Psi^{jc}\big) -\epsilon^{ij} \big(\sigma_k^a  \Psi^{kb} +\sigma_k^b  \Psi^{ka} \big)\right), \label{formS1} \\
-\epsilon^{il} D^{kc} \big( \cS_J \big)^{ad} -\epsilon^{ik} D^{ld} \big( \cS_J \big)^{ac} - \epsilon^{ad}D^{kc}\big( \cS_T \big)^{il}  -\epsilon^{ac}D^{ld}\big( \cS_T \big)^{ik} -2\im \epsilon^{kl}\epsilon^{cd}\Psi^{ia}  =\nn \\
=-\im \epsilon^{il} \epsilon^{ad}\Psi^{kc} -\im \epsilon^{ik}\epsilon^{ac} \Psi^{ld}
+ \frac{1}{3} \big( -\epsilon^{ac} \epsilon^{il} \sigma^{kb}\big( \cS_J \big)_b{}^d - \epsilon^{ad}\epsilon^{ik}\sigma^{lb}\big( \cS_J \big)_b{}^c  +\nn \\
+ \epsilon^{il}\sigma^{ka}\big( \cS_J \big)^{cd} + \epsilon^{ik}\sigma^{la}\big( \cS_J \big)^{cd} + \epsilon^{ik}\sigma^{lc}\big( \cS_J \big)^{ad} + \epsilon^{il}\sigma^{kd}\big( \cS_J \big)^{ac} + \epsilon^{kl}\epsilon^{cd} \sigma^{ib} \big( \cS_J \big)_b{}^a \big) + \nn \\ + \frac{1}{3}\big( - \epsilon^{ac}\sigma^{kd}\big(  \cS_T \big)^{il} - \epsilon^{ad}\sigma^{lc}\big( \cS_T  \big)^{ik} -\epsilon^{ad} \sigma^{ic}\big( \cS_T  \big)^{kl} - \epsilon^{ac}\sigma^{id} \big( \cS_T \big)^{kl}    + \nn \\+
\epsilon^{kl}\epsilon^{cd} \sigma^a_m \big( \cS_T \big)^{im} -\epsilon^{ik}\epsilon^{ad}\sigma^c_m \big( \cS_T \big)^{lm} - \epsilon^{il}\epsilon^{ac}\sigma^d_m \big( \cS_T \big)^{km}  \big) . \label{formS2}
\eea
Equation \p{formS2} is satisfied identically after substitution of $\Psi^{ia}$ \p{Psi} and $\big(\cS_J \big)^{ab}$, $\big( \cS_T  \big)^{ij}$ \p{SJST}. In this calculation, one should use the formula
\be\label{DDF}
D^{jb}D^a_k F^{ik} = 3\im \epsilon^{ab}{\dot F}{}^{ij} - \frac{1}{4}\epsilon^{ij}\epsilon^{ab}D^c_k D_{cl}F^{kl},
\ee
which follows from \p{DFFF} and commutation relations \p{covders}. Equation \p{formS1} after substitution  of $\Psi^{ia}$ \p{Psi} and $\big(\cS_J \big)^{ab}$, $\big( \cS_T  \big)^{ij}$ \p{SJST} reduces to $\epsilon^{ij}\epsilon^{ab}$-projection, which determines $C$
\bea\label{C}
\alpha (1+\alpha)C &=& - \frac{\im}{3}\alpha {\dot F} + \frac{1}{72}D^c_k D_{lc}F^{kl} + \frac{1}{9}F^2 - \frac{1}{72\alpha}F_{kl}F^{kl} + \frac{1}{72(1+\alpha)} F_{cd}F^{cd} - \nn \\
&&- \frac{1}{54(1+\alpha)} F^{cd}\sigma_{mc}\sigma^m_d -  \frac{1}{54\alpha} F^{kl}\sigma_{kc}\sigma^c_l + \frac{\im}{9}{\dot\sigma}{}^{kc}\sigma_{kc} + \frac{1}{162}\left( \frac{1}{\alpha} + \frac{1}{1+\alpha}  \right) \sigma_{kc}\sigma^k_d \sigma^c_l \sigma^{ld}.
\eea

Finally, studying  $d\omega_K$ equation we will get the following relations
\bea
&& \im {\dot \Psi}{}^{ia}- D^{ia} C = -2\im \left( \Psi^i_c  \left(\cS_J\right)^{ca}+\Psi^a_k  \left(\cS_T\right)^{ki}\right), \label{omeKa} \\
&&2 \epsilon^{ij} \epsilon^{ab} C - D^{ia} \Psi^{jb} -D^{jb} \Psi^{ia} = \epsilon^{ij}\epsilon^{ab} \left( \cS_J^2 +\cS_T^2 \right) - 4 \big( \cS_T\big){}^{ij} \big( \cS_J \big){}^{ab} .  \label{omeKb}
\eea
Here, $\cS_J^2 \equiv (\cS_J)^{ab} (\cS_J)_{ab}$  and  $\cS_T^2 \equiv (\cS_T)^{ij} (\cS_T)_{ij}$.
The relation \p{omeKb} is simply a consequence of \p{formS1} and therefore, is satisfied identically. Finally, substituting $C$ \p{C}, $\Psi^{ia}$ \p{Psi} and $\big(\cS_J \big)^{ab}$, $\big( \cS_T  \big)^{ij}$ \p{SJST} into \p{omeKa} and using formula
\be\label{DDDF}
D^{ia}D^c_k D_{lc} F^{kl} = - 4\im D^a_l {\dot F}{}^{il},
\ee
one finds that \p{omeKa} is also identically satisfied, leaving no extra constraints on $\sigma^{ia}$.

\end{document}